\documentclass[a4paper,11pt]{article}
\usepackage{pos}

\usepackage{graphicx} 
\usepackage{float}

\usepackage[numbers,sort&compress]{natbib}

\usepackage{tensor} 

\def\nn{\nonumber}

\title{Characters, Quasinormal Modes, and Quantum de Sitter Thermodynamics}

\author[a,b]{Y. T. Albert Law}

\affiliation[a]{ \it \normalsize	 Center for the Fundamental Laws of Nature, Harvard University, Cambridge, MA 02138, USA}
\affiliation[b]{ \it \normalsize	 Black Hole Initiative, Harvard University, Cambridge, MA 02138, USA}

\emailAdd{ylaw1@g.harvard.edu}

\abstract{In this short note, we review some recent progress in understanding the 1-loop corrections to the Gibbons-Hawking entropy, which amounts to studying free fields on the de Sitter static patch and the round sphere. After briefly surveying the unitary irreducible representations of the de Sitter group $SO(1,d+1)$ and their Harish-Chandra characters, we discuss the Lorentzian interpretation for the 1-loop sphere path integral for a scalar. After that we comment on how the results are modified by edge contributions for spinning fields.}

\FullConference{%
  Corfu Summer Institute 2022 "School and Workshops on Elementary Particle Physics and Gravity",\\
  28 August - 1 October, 2022\\
  Corfu, Greece}

\begin{document}
\maketitle


\tableofcontents


\section{Introduction}

Due to the exponential cosmic expansion, the causal diamond for an inertial observer in a de Sitter (dS) spacetime $dS_{d+1}$ covers only a part of it known as the static patch, described by the metric
\begin{align}\label{eq:staticmetric}
	ds^2 = -\left( 1-\frac{r^2}{\ell_\text{dS}^2}\right) dt^2 + \frac{dr^2}{ 1-\frac{r^2}{\ell_\text{dS}^2}}+r^2 d\Omega_{S^{d-1}}^2 \; .
\end{align}
The dS length $\ell_\text{dS}$ is related to the cosmological constant $\Lambda>0$ through $\ell_\text{dS} = \sqrt{\frac{d(d-1)}{2\Lambda}}$. Sitting at $r=0$, the observer is surrounded by a cosmological horizon of area $A_c = \ell_\text{dS}^{d-1}\text{Vol}(S^{d-1})$ at $r=\ell_\text{dS}$. A semiclassical analysis tells us that this horizon has a Hawking temperature $T_\text{dS}=\frac{1}{2\pi \ell_\text{dS}}$ and an associated Gibbons-Hawking entropy \cite{PhysRevD.15.2738} (from now on we set $\ell_\text{dS}=1$)
\begin{align}\label{eq:GHentropy}
	S_\text{GH}=\frac{A_c}{4G_N} \;, 
\end{align}
whose microscopic origin remains a mystery. 

In the absence of a microscopic explanation, a bottom-up approach would to be study the quantum corrections to  \eqref{eq:GHentropy} in the low energy effective theory of gravity plus matter. These corrections provide unambiguous data testing candidate models, in similar spirit of that in the context of black hole microstate counting \cite{Banerjee:2010qc, Banerjee:2011jp, Sen:2012dw, Sen:2012kpz, Sen:2014aja} and Higher-spin/CFT dualities \cite{Giombi:2013fka,Giombi:2014iua,Giombi:2016pvg,Gunaydin:2016amv}. As an illustration, let us say one asserts that the dS horizon entropy $\mathcal{S}_\text{macro}$ for 3D pure gravity (computed in \cite{Anninos:2020hfj}) counts the number $W(N)$ of partitions of $N$.\footnote{For large $N$, $W(N)\approx \frac{1}{4N\sqrt{3}}e^{\pi\sqrt{2N/3}}$.} Both the macroscopic and microscopic entropies can be brought uniquely into an expansion of the form
\begin{align}
	\mathcal{S}_\text{macro} =& \mathcal{S}_0 -3 \log \mathcal{S}_0 + \cdots \nn\\
	\mathcal{S}_\text{micro} \equiv & \log W(N) =  \mathcal{S}_0 -2 \log \mathcal{S}_0 + \cdots \; .
\end{align}
The failure to reproduce $\mathcal{S}_\text{macro}$ immediately falsifies the asserted microscopic proposal. In contrast, as shown in \cite{Shyam:2021ciy,Coleman:2021nor}, a $T \bar T(+\Lambda_2)$-deformed $\text{CFT}_2$ reproduces $\mathcal{S}_\text{macro}$ for 3D pure gravity up to the logarithmic correction. 

This note reports some recent progress \cite{Anninos:2020hfj,Law:2020cpj,Law:2022zdq} in studying the 1-loop corrections to \eqref{eq:GHentropy}, which as reviewed in Section \ref{sec:sphereLor} comes from the quadratic fluctuations of gravitons and matter in the Euclidean path integral around the round $S^{d+1}$ saddle. These contribute to the full dS entropy as
\begin{align}\label{eq:dsentropyex}
	\mathcal{S}_\text{macro} = \cdots + \underbrace{b \log \mathcal{S}_0 +c}_{\text{1-loop}} +\cdots \; .
\end{align}
The logarithmic correction (i.e. the coefficient $b$) depends only on the massless spectrum while the finite part $c$ receives contributions from the full matter content. 

In understanding the 1-loop corrections, a key role is played by the Harish-Chandra character of the de Sitter group $SO(1,d+1)$. Our goals in this note are two-fold: first, to explain how these mathematically defined objects lead to a {\it Lorentzian} interpretation of the 1-loop Euclidean path integral; second, to describe qualitatively new features as one extends the discussion to spinning fields. We will restrict ourselves to generic dimensions $d\geq 3$ to avoid discussing special features that only appear in $d< 3$.

After setting up the problem in Section \ref{sec:sphereLor}, we will do a quick survey for unitary irreducible representations (UIRs) of $SO(1,d+1)$ in Section \ref{sec:dSUIR}. In Section \ref{sec:char}, we introduce the $SO(1,d+1)$ Harish-Chandra character and explain its physics. We discuss in Section \ref{sec:PI} generalizations to spinning fields, before we conclude in Section \ref{sec:remarks}. As we aim to keep the discussion concise, many details for the derivations (including for example a careful treatment of UV-divergences) will be inevitably omitted. We refer the interested readers to the original work \cite{Anninos:2020hfj,Law:2020cpj,Law:2022zdq} for the complete detail.


\section{What is the Lorentzian interpretation of the 1-loop sphere path integral?}\label{sec:sphereLor}

Proposed by Gibbons and Hawking \cite{PhysRevD.15.2752}, the full dS entropy is computed by a Euclidean path integral with a positive cosmological constant $\Lambda$, integrating over all metrics $g$ and matter fields collectively denoted by $\Phi$:
\begin{align}\label{eq:fulldSentropy}
	\mathcal{S} = \log \mathcal{Z}\; , \qquad \mathcal{Z} = \int \mathcal{D} g\, \mathcal{D}\Phi \, e^{-S[g,\Phi]}  \; ,
\end{align}
expanded around the round sphere $S^{d+1}$ saddle. At tree level in Einstein gravity, one recovers the Gibbons-Hawking entropy \eqref{eq:GHentropy}. At 1-loop, one integrates the quadratic fluctuations around this saddle, which for a scalar is given by the functional determinant of a Laplace operator on the sphere (whose radius is set to 1)
\begin{align}\label{eq:scalarPI}
	Z_\text{PI} = \int \mathcal{D}\phi \, e^{-\frac{1}{2}\int_{S^{d+1}}\phi\left(-\nabla^2+m^2\right) \phi} = \det \left( -\nabla^2+m^2\right)^{1/2} \; .
\end{align}
Such a quantity is UV-divergent and is typically regulated by for example zeta-function \cite{Hawking:1976ja} or heat kernel \cite{Vassilevich:2003xt} regularization.

\begin{figure}[H]
	\centering
	\includegraphics[scale=0.3]{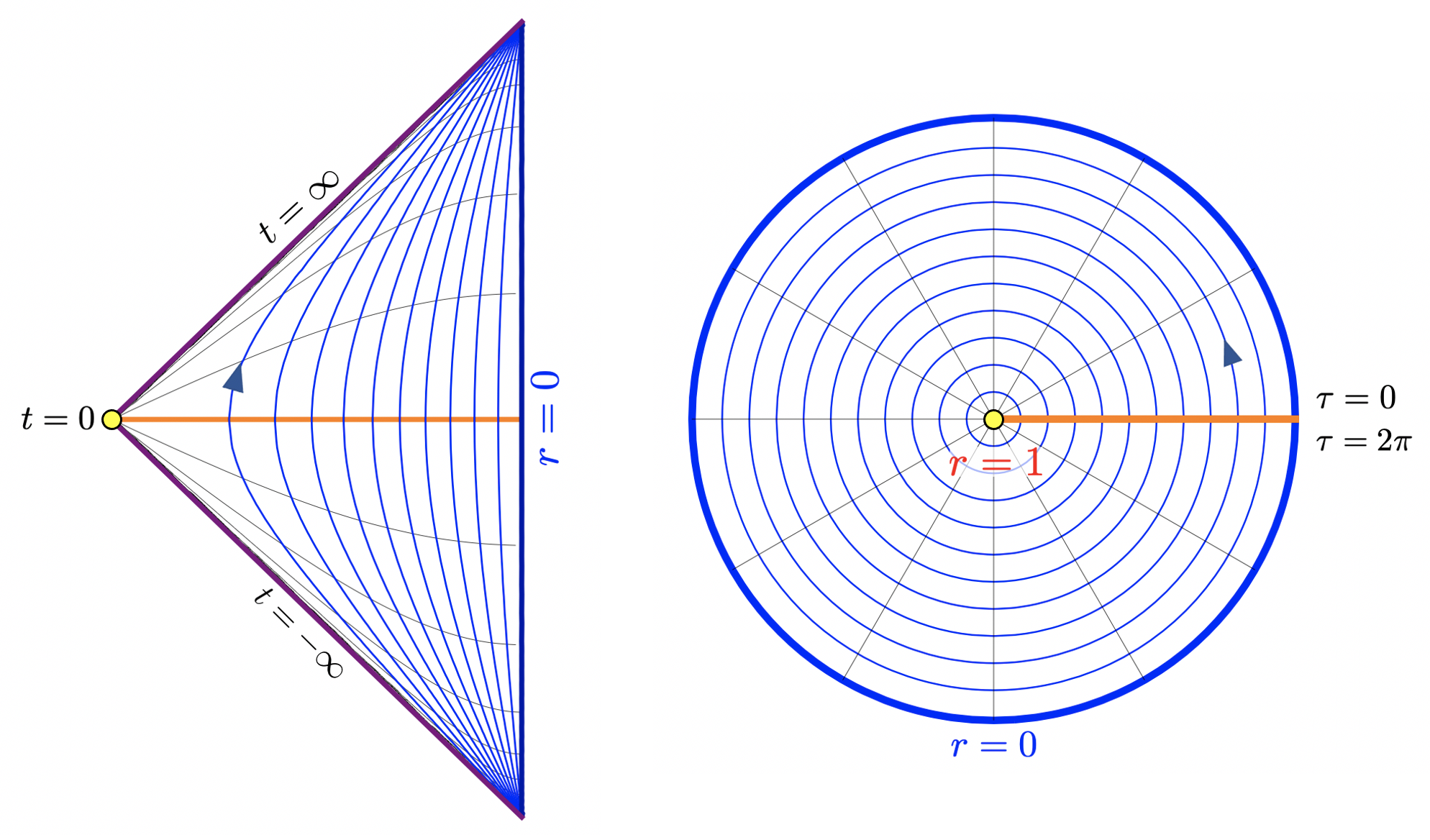}
	\caption{After Wick-rotating $t \to -i \tau$ and identifying $\tau \sim \tau +2\pi $, the dS static patch (left) turns into a sphere (right). The Euclidean time $\tau$ becomes an angular variable on the sphere. The horizon (yellow dot) at $r=1$ is mapped to the origin of the sphere. }
	\label{fig:dSsphere}
\end{figure}

Since $S^{d+1}$ can be obtained by Wick-rotating the static patch time $t\to - i \tau$ and identifying $\tau \simeq \tau + 2\pi$ in \eqref{eq:staticmetric} (see Fig.\ref{fig:dSsphere}), one expects that a 1-loop path integral on $S^{d+1}$ would have an interpretation as a thermal ideal gas canonical partition function for quantum fields living on a static patch at the dS temperature $T_\text{dS}=\frac{1}{2\pi}$. For a free massive scalar, this means
\begin{align}\label{eq:PItrace}
	Z_\text{PI} \sim Z_\text{bulk} \equiv \text{``Tr"} \, e^{-2\pi \hat{H}}
\end{align}
where $\hat{H}$ is the static patch Hamiltonian with respect to which one defines positive energy modes, and ``Tr"" is tracing over the associated Fock space. We use $\sim$ instead of an equal sign and put a quotation mark on Tr because the trace is in fact ill-defined. To see this, let us pretend that the spectrum of $\hat{H}$ were discrete. One could then compute the trace \eqref{eq:PItrace} by substituting the mode expansion for the scalar field and summing over occupation numbers, leading to 
\begin{align}\label{intro:Zbulkdis}
	Z_\text{bulk}= \prod_{E>0} \frac{e^{-\pi E}}{1-e^{-2\pi E}} \; .
\end{align}
Here the product is over the discrete {\it single-particle} energy spectrum labeled by $E$; the factor $e^{-\pi E}$ is due to the zero-point energy for each positive-energy mode. Equivalently, we can write
\begin{align}\label{intro:log Z bulk}
	\log Z_\text{bulk} =-\int_0^\infty d\omega \, \rho(\omega) \log \left( e^{\pi\omega}-e^{-\pi\omega}\right) 
\end{align}
in terms of the single-particle density of states (DOS)
\begin{align}\label{intro:dosdis}
	\rho(\omega) = \sum_{E>0} \delta(\omega -E) = \text{tr} \, \delta(\omega -\hat H) 
\end{align}
with tr tracing over the single-particle Hilbert space. 

The trouble for us is that the spectrum of $\hat H$ is actually continuous, and the DOS $\rho(\omega)$ is strictly infinite. Physically, this is due to the fact that the horizon is an infinite redshift surface, enabling the existence of normal modes with arbitrary angular momentum and energy. Subsequently, expressions \eqref{intro:Zbulkdis}-\eqref{intro:dosdis} do not make sense, casting doubts on the interpretation \eqref{eq:PItrace}. What comes to rescue is the Harish-Chandra character for the de Sitter group $SO(1,d+1)$, which enables us to make sense of the right hand side of \eqref{eq:PItrace}. Before explaining that, we will first quickly review the symmetries of $dS_{d+1}$ and the UIRs of the de Sitter group $SO(1,d+1)$.



\section{Unitary irreducible representations of the de Sitter group $SO(1,d+1)$}\label{sec:dSUIR}

\subsection{The geometry of $dS_{d+1}$ and its symmetries}

The $(d+1)$-dimensional dS space $dS_{d+1}$  can be thought of as a hyperboloid in $\mathbb{R}^{1,d+1}$
\begin{align}\label{eq:dSembed}
	\eta_{AB}X^A X^B = -\left( X^0\right)^2 +\left( X^1\right)^2 +\cdots +\left( X^{d+1}\right)^2= 1 \; .
\end{align}
The hyperbloid \eqref{eq:dSembed} has a manifest $SO(1,d+1)$ isometry generated by Killing vectors $L_{AB} = X_A \partial_{X^B}- X_B \partial_{X^A}$, which satisfies the algebra
\begin{align}\label{eq:dSalgebra}
	\left[ L_{AB},L_{CD}\right]= \eta_{BC}L_{AD}- \eta_{AC}L_{BD}+ \eta_{AD}L_{BC}- \eta_{BD}L_{AC}\; .
\end{align}
Defining 
\begin{align}\label{eq:congen}
	D=L_{0,d+1}\; , \quad M_{ij}=L_{ij}\; , \quad P_i = L_{d+1,i}+L_{0,i}\; , \quad K_i = L_{d+1,i}-L_{0,i} \; ,
\end{align}
the algebra \eqref{eq:dSalgebra} is recast into the conformal algebra of $\mathbb{R}^d$:
\begin{gather}
	[D,P_i]=P_i \; , \qquad [D,K_i]=-K_i \; , \qquad [K_i,P_j]=2\delta_{ij} D-2 M_{ij} \;,\nn\\
	[M_{ij},P_k] = \delta_{jk} P_i - \delta_{ik} P_j \; ,\qquad [M_{ij},K_k] = \delta_{jk} K_i - \delta_{ik} K_j \; .
\end{gather}
The static patch \eqref{eq:staticmetric} corresponds to the intrinsic parametrization 
	\begin{align}
		X^0 =\sqrt{1-r^2} \sinh t \; , \qquad X^{d+1}= \sqrt{1-r^2} \cosh t \; , \qquad X^i = r \,  \Omega_{S^{d-1}}^i \;,
	\end{align}
and has a manifest $SO(1,1)\times SO(d)$ symmetry corresponding to $t$-translation and rotations on $S^{d-1}$ generated by the boost $D$ and angular momentum $M_{ij}$ respectively. Other isometries of the global dS space, i.e. $P_i$ or $K_i$, do not preserve a single static patch but instead map different static patches into one another.

\subsection{Particles in $dS_{d+1}$= UIRs of $SO(1,d+1)$}

Analogous to Wigner's classification of particles and fields in Minkowski spacetime by unitary irreducible representations (UIRs) of the Poincar\'{e} group, particles and fields in $dS_{d+1}$ are classified according to UIRs of the de Sitter group $SO(1,d+1)$. For an excellent review of the representation theory of $SO(1,d+1)$, we refer the reader to \cite{Sun:2021thf}. Here we will only do a quick survey for the dictionary between UIRs of $SO(1, d+1)$ and QFTs in $dS_{d+1}$. 

A $SO(1,d+1)$ UIR $\mathcal{V}_{[\Delta, \mathbf{s}]}$ is labeled by an $SO(d)$ highest weight $\mathbf{s}=[s_1,s_2,\dots,s_{\lfloor \frac{d}{2}\rfloor}]$ and a conformal dimension $\Delta$; the former is nothing but the spin for the quantum field, while the latter is related to its mass. We focus on scalars and symmetric tensors, i.e. those with $\mathbf{s}=[s,0,\dots,0]$ and $s\geq 0$ is an integer. For a spin-$s$ field, its mass is related to the conformal dimension through 
\begin{align}\label{eq:massdim}
	m^2 =
	\begin{cases}
		\Delta (d-\Delta)\; , & s=0 \\
		(\Delta+s-2)(d+s-2-\Delta) \;, & s\geq 1 \\
	\end{cases}\;.
\end{align}
Unitarity requires that the generators \eqref{eq:dSalgebra} to act as {\it anti-hermitian} operators on the UIR $\mathcal{V}_{[\Delta, \mathbf{s}]}$, i.e.
\begin{align}\label{eq:unicon}
	L_{AB}^\dagger = -L_{AB} \; . 
\end{align}
In terms of the conformal generators \eqref{eq:congen}, these imply
\begin{align}\label{eq:uniconcon}
	D^\dagger = -D\; , \qquad P_i^\dagger=-P_i \;, \qquad K_i^\dagger=-K_i \; , \qquad M_{ij}^\dagger = -M_{ij}\;.
\end{align}
These conditions will restrict the form of the inner products (with respect to which one defines the hermitian conjugation) and the allowed values of $\Delta$. Note that \eqref{eq:uniconcon} is distinct from the unitarity condition of the {\it Lorentzian} conformal group $SO(2,d)$ which demands for example $P_i^\dagger = K_i$, marking a fundamental difference between QFTs in de Sitter and anti-de Sitter space.


\paragraph{Principal series representations}

These UIRs describe massive fields in $dS_{d+1}$ that are heavy compared with the dS scale $\ell_\text{dS}$. For a spin-$s$ field, its conformal dimension and mass fall into the ranges
\begin{align}
	\Delta = \frac{d}{2}+ i\nu \; , \quad \nu \in \mathbb{R} \quad \Leftrightarrow \quad m\geq 
	\begin{cases}
		\frac{d}{2} &\text{ for } s=0 \\
		 s+\frac{d}{2}-2 &\text{ for } s\geq 1
	\end{cases}\;.
\end{align}

\paragraph{Complementary series representations}

These UIRs describe massive fields in $dS_{d+1}$ that are light compared with the dS scale $\ell_\text{dS}$. For a light scalar,  it has
\begin{align}
	0 < \Delta <d \qquad \Leftrightarrow  \qquad 0<m<\frac{d}{2} \qquad  (s=0) \; .
\end{align}
For a light spin-$s$ field, the ranges are instead
\begin{align}
	1<\Delta <d-1 \qquad \Leftrightarrow \qquad  \sqrt{(s-1)(s+d-3)}<m< s+\frac{d}{2}-2 \qquad (s\geq 1) \; .
\end{align}
The lower bound $\sqrt{(s-1)(s+d-3)}$ is known as the Higuchi bound \cite{Higuchi:1986py}.

\paragraph{Exceptional series}

These UIRs only exist for $s\geq 1$ and describe the so-called partially massless particles \cite{Higuchi:1986py,Zinoviev:2001dt,Deser:1983tm,DESER1984396,Brink:2000ag,Deser:2001pe,Deser:2001us, Deser:2001wx,Deser:2001xr, Skvortsov:2006at, Hinterbichler:2016fgl,Basile:2016aen}.\footnote{In the terminology of \cite{Sun:2021thf}, these UIRs are ``exceptional series II". ``Exceptional series I" might describe the so-called shift-symmetric scalars \cite{Bonifacio:2018zex}, with the shift symmetry gauged.} These occur when the conformal dimension and mass hit the exceptional points
\begin{align}
	\Delta = 1- p \qquad \Leftrightarrow \qquad m_{s,p}^2=(s-1-p)(d+s+p-3) \; , \qquad  p=0,1,\dots,s-1 \; ,
\end{align}
The local action describing these fields have the gauge symmetry that reads schematically
\begin{align}
	\phi_{\mu_1 \cdots \mu_s} \sim \phi_{\mu_1 \cdots \mu_s} + \nabla_{(\mu_{s-p}} \cdots \nabla_{\mu_s} \xi_{\mu_1 \cdots \mu_p)} + \dots \; ,
\end{align}
where $\cdots$ stand for terms with fewer derivatives \cite{Hinterbichler:2016fgl}. We call the integer $p$ "depth", which is equal to the spin of the gauge parameter. In particular, the usual massless spin-$s$ field has depth $p=s-1$.


\section{Harish-Chandra character of $SO(1,d+1)$ and its physics}\label{sec:char}

A useful object to encode the information about a $SO(1,d+1)$  UIR is the Harish-Chandra character, defined as a trace of a group element over the representation space $\mathcal{V}_{[\Delta, \mathbf{s}]}$. The general theory for Harish-Chandra characters and their computations in the case of $SO(1,d+1)$ are reviewed in \cite{Sun:2021thf}. For our purpose, we will focus on the character for the group element generating time translation in the static patch
\begin{align}\label{eq:redchar}
	\chi_{[\Delta, \mathbf{s}]}(t) = \text{tr}_{[\Delta, \mathbf{s}]} \, e^{- i\hat{H} t} \; .
\end{align}
We have written in terms of $\hat{H}=i \hat{D}$, which can be interpreted as a {\it hermitian} Hamiltonian with a real spectrum. The character \eqref{eq:redchar} has the property that $\chi_{[\Delta, \mathbf{s}]}(-t)=\chi_{[\Delta, \mathbf{s}]}(t)$. In the following discussion, we will restrict to $t>0$ unless specified.

\paragraph{Principal and complementary series}

For these UIRs, the characters have a very simple form: 
\begin{align}\label{eq:masschar}
	\chi_{[\Delta,s]}(t) = D_s^d\frac{e^{-\Delta t}+e^{-\bar\Delta t}}{(1-e^{-t})^d} \; , \qquad \bar \Delta \equiv d-\Delta\; ,
\end{align}
where we recall the relation \eqref{eq:massdim} between the mass and conformal dimension. The number 
\begin{align}
	D_s^d = \frac{2s+d-2}{d-2}\binom{s+d-3}{s}
\end{align}
is the dimension of the spin-$s$ representation of $SO(d)$, or the number of independent polarizations of a spin-$s$ field.

\paragraph{Exceptional series}

Due to the presence of gauge invariance, the construction of exceptional series representations are more intricate than principal or complementary series. This intricacy is also reflected in their characters. For example, a massless spin-1 field has the character
\begin{align}\label{eq:masslessvecchar}
	\chi_{[\Delta=1,s=1]}(t) = 1-\frac{1- d e^{- t}}{(1-e^{-t})^d}+\frac{d e^{-(d-1) t}-e^{-d t}}{(1-e^{-t})^d} \; .
\end{align}

\subsection{Quasinormal mode expansion}

Recall that a quasinormal mode (QNM) on a static patch satisfies two boundary conditions \cite{Brady:1999wd,Lopez-Ortega:2006aal,Lopez-Ortega:2012xvr}: it is regular at the location of the observer and is purely outgoing into the horizon. Such conditions force the energies to take discrete and in general complex values, as opposed to {\it normal} modes whose energies are real and continuous. As pointed out in \cite{Anninos:2020hfj,Sun:2020sgn}, the Harish-Chandra character encodes QNMs on a static patch. Explicitly, expanding in powers of $e^{-t}$, a general character takes the form
\begin{align}
	\chi_{[\Delta, \mathbf{s}]} (t) = \sum_z N_z \, e^{-i z t}
\end{align}
where $z$ denotes a QNM frequency with degeneracy $N_z$. For instance, the QNM expansion for the massive scalar character reads
\begin{align}
	\chi_{[\Delta,s=0]}(t) =& \frac{e^{-\Delta t}+e^{-\bar\Delta t}}{(1-e^{-t})^d}=\sum_{n=0}^\infty\sum_{l=0}^\infty D_l^d \,  \left[e^{-i 	z_{\Delta,n,l}  t}+e^{-iz_{\bar \Delta,n,l}t}\right] \; ,
\end{align}
where 
\begin{align}\label{eq:qnmfreq}
	i z_{\Delta,n,l} = \Delta+2n+l \; , \qquad	i z_{\bar \Delta,n,l} = \bar\Delta+2n+l
\end{align}
are the QNM frequencies labeled by the overtone number $n$ and the $SO(d)$ quantum number $l$, with degeneracy $D_l^d=\frac{2l+d-2}{d-2}\binom{l+d-3}{d-3}$. The underlying representation theory enables an algebraic construction of QNMs in dS space, originally pointed out in \cite{Ng:2012xp,Jafferis:2013qia,Tanhayi:2014kba} and extended to the (massive and massless) higher spin fields in \cite{Sun:2020sgn}.

\subsection{Spectral density and the quasicanonical partition function}\label{sec:qnmPF}

The key observation in \cite{Anninos:2020hfj} is that the Fourier-transform of the character \eqref{eq:redchar} can be interpreted as a spectral density  of the Hamiltonian $\hat{H}$:
\begin{align}\label{eq:chardos}
	\tilde{\rho}_{[\Delta, \mathbf{s}]}(\omega) \equiv \int_{-\infty}^\infty \frac{dt}{2\pi}e^{i\omega t} \chi_{[\Delta, \mathbf{s}]}(t)= \int_{0}^\infty \frac{dt}{2\pi}\left( e^{i\omega t}+e^{-i\omega t} \right) \chi_{[\Delta, \mathbf{s}]}(t) = \text{tr}_{[\Delta, \mathbf{s}]} \, \delta(\omega -\hat{H}) \; ,
\end{align} 
where in the second step we recall $\chi_{[\Delta, \mathbf{s}]}(-t)=\chi_{[\Delta, \mathbf{s}]}(t)$. One should keep in mind that the trace tr is over the {\it representation space} $\mathcal{V}_{[\Delta, \mathbf{s}]}$, which is distinct from the single-particle trace \eqref{intro:dosdis} over the Fock space in the static patch. Nonetheless, if one {\it replaces} the ill-defined DOS $\rho(\omega)$ in \eqref{intro:log Z bulk} by \eqref{eq:chardos}, one could define a quasicanonical or renormalized partition function \cite{Anninos:2020hfj}
\begin{align}\label{eq:canPF}
	\log \widetilde{Z}_\text{bulk} \equiv  \int_0^\infty \frac{dt}{2t} \frac{1+e^{- t}}{1-e^{- t}}\,\chi_{[\Delta, \mathbf{s}]}(t) \; .
\end{align}
Here we have performed the $\omega$-integral. What is more, \cite{Anninos:2020hfj} finds that for scalars this is precisely equal to the 1-loop sphere path integral \eqref{eq:scalarPI}, i.e.\footnote{Note that either \eqref{eq:scalarPI} or \eqref{eq:canPF} are UV-divergent and need to be regulated. The equality \eqref{eq:PIPFs0} means that the scheme-independent part (e.g. the coefficient of the logarithmic divergence) of both sides agree.}
\begin{align}\label{eq:PIPFs0}
	Z_\text{PI} = \widetilde{Z}_\text{bulk} \qquad \text{(Scalar)}
\end{align}
while for spinning fields the 1-loop sphere path integrals $Z_\text{PI}$ are given by \eqref{eq:canPF} with ``edge" corrections, which we will further discuss in section \ref{sec:PI}.

\subsection{Horizon scattering}\label{sec:scattering}

\begin{figure}[H]
	\centering
	\includegraphics[scale=0.4]{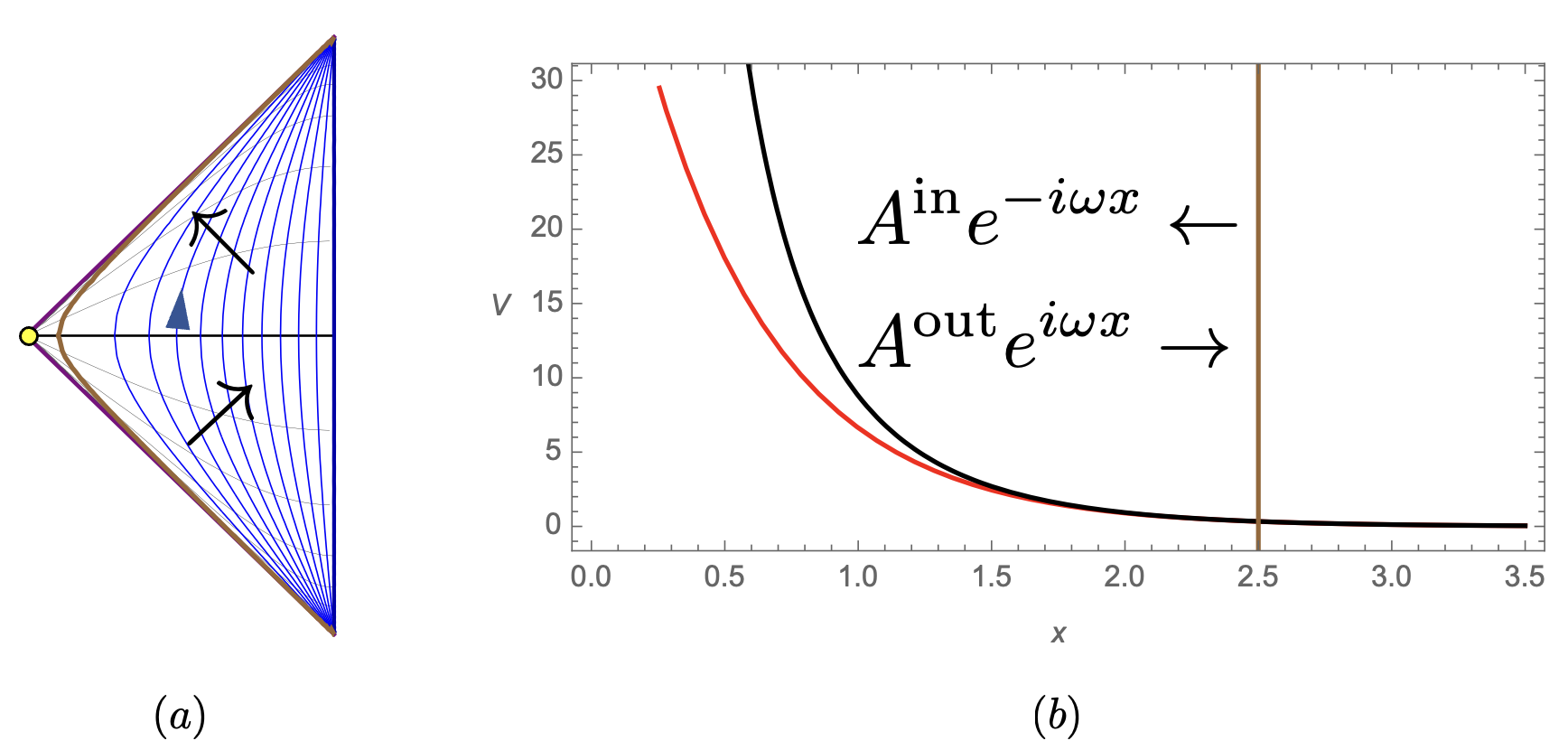}
	\caption{(a) The Penrose diagram for a static patch (b) The scattering potential \eqref{eq:scatpot} (black) for a scalar with $\Delta=1.5+ 0.1 i $ and $l=3$ on a static patch in on $dS_4$ , which is hardly distinguishable from the Rindler potential (red) in \eqref{eq:Rindlerscatter} for $x\gg 0$. All quantities are in units of the dS length $\ell_\text{dS}$ or its inverse. The horizon is at $x=\infty$. In these diagrams, we indicate the incoming and outgoing waves \eqref{eq:scattasym} with black arrows. The brown lines indicate the brick wall regulator \cite{tHooft:1984kcu} at $x=R=2.5\ell_\text{dS}$, corresponding to $r\approx 0.987\ell_\text{dS}$.}
	\label{fig:scattering}
\end{figure}

A physical picture of the spectral density \eqref{eq:chardos} and the resulting partition function \eqref{eq:canPF} from the perspective of a single static patch is elaborated (and extended to general static black holes) in \cite{Law:2022zdq}. We use a scalar as an example. For each real energy $\omega>0$ and $SO(d)$ angular momentum $l\geq 0$, a normal mode takes the form
\begin{align}\label{eq:normalmodes}
	\phi_{\omega l} (t,r,\Omega)= e^{-i \omega t}\,\frac{\psi(r)}{r^\frac{d-1}{2} } \, Y_l(\Omega)\; . 
\end{align}
Here $Y_l$ are the $(d-1)$-dimensional spherical harmonics with degeneracy $D_l^d$ satisfying $-\nabla^2_{S^{d-1}}Y_l=l(l+d-2)Y_l$. We require the modes \eqref{eq:normalmodes} to be regular at the location of the observer. With \eqref{eq:normalmodes}, the Klein-Gordon equation $\left( -\nabla^2+m^2\right) \phi =0$ is equivalent to a scattering problem
\begin{align}\label{eq:eff Sch}
	\left(-\partial_x^2 +V_l(x) \right) \psi(x)=\omega^2 \psi(x) \; , \qquad 0\leq x <\infty\; ,
\end{align}
where $ x =\ell_\text{dS}\tanh^{-1}\frac{r}{\ell_\text{dS}} $ is the tortoise coordinate (we restore the dS length $\ell_\text{dS}$ for the rest of this section). The scattering potential reads explicitly
\begin{align}\label{eq:scatpot}
	V_l(x) = -\frac{(d-2 \Delta -1) (d-2 \Delta +1)}{4\ell_\text{dS}^2\cosh^2 \frac{x}{\ell_\text{dS}}} +\frac{(d+2 l-3) (d+2 l-1)}{4\ell_\text{dS}^2\sinh^2 \frac{x}{\ell_\text{dS}}}  \;.
\end{align}
Therefore, normal modes \eqref{eq:normalmodes} are equivalent to scattering modes for the scattering problem \eqref{eq:eff Sch}: they describe plane waves come in from the horizon at $x=\infty$ and get scattered off the potential \eqref{eq:scatpot} centered at the observer (at $x=0$) before going back into the horizon. See Fig. \ref{fig:scattering}.
Concretely, the solution to \eqref{eq:eff Sch} that is regular at $x=0$ is a mixture of plane waves near horizon
\begin{gather}\label{eq:scattasym}
	\psi_l (x\to \infty ) \sim A_l^\text{out}(\omega)\, e^{i\omega x} + A_l^\text{in}(\omega) \, e^{-i\omega x} \; .
\end{gather}
Here for real $\omega$ the incoming and outgoing coefficients 
\begin{align}\label{eq:inoutco}
	A_l^\text{out}(\omega) = \left( A_l^\text{in}(\omega)\right)^* =\frac{\Gamma\left(i \ell_\text{dS} \omega\right)}{\Gamma\left(\frac{\Delta+l+i \ell_\text{dS} \omega}{2}\right) \Gamma\left(\frac{\bar\Delta+l+ i \ell_\text{dS} \omega}{2}\right)} 
\end{align}
are complex conjugates to each other and thus the ratio $\mathcal{S}_l (\omega) \equiv A_l^\text{out}(\omega)/A_l^\text{in}(\omega)$ is a pure phase. In fact, from \eqref{eq:inoutco} we see that $\mathcal{S}_l (\omega) $ is a product of two phases:
\begin{align}\label{eq:dSSmat}
	\mathcal{S}_l (\omega) = \mathcal{S}^\text{dS}_l (\omega) \mathcal{S}^\text{Rindler} (\omega)
\end{align}
where
\begin{align}\label{eq:dSdress}
	\mathcal{S}_l^\text{dS} (\omega)\equiv \frac{\Gamma\left(\frac{\Delta+l- i \ell_\text{dS} \omega}{2}\right) \Gamma\left(\frac{\bar\Delta+l- i \ell_\text{dS} \omega}{2}\right)}{\Gamma\left(\frac{\Delta+l+i \ell_\text{dS} \omega}{2}\right) \Gamma\left(\frac{\bar\Delta+l+ i \ell_\text{dS} \omega}{2}\right)} 
\end{align}
has the QNM frequencies \eqref{eq:qnmfreq} as poles and their complex conjugates as zeros. In other words, QNMs are scattering resonances associated with the scattering problem \eqref{eq:eff Sch}. The other phase 
\begin{align}\label{eq:RindSmat}
	 \mathcal{S}^\text{Rindler} (\omega) = \frac{\Gamma\left(i \ell_\text{dS} \omega\right)}{\Gamma\left(-i \ell_\text{dS} \omega\right)}
\end{align}
is $l$-independent and captures all the Matsubara frequencies
\begin{align}\label{eq:Matfreq}
\omega = \omega_{\pm , n} = \pm i \frac{n}{ \ell_\text{dS}} \; , \qquad n = 1 , 2 , 3 ,\cdots  \; ,
\end{align}
as its poles and zeros. 

\paragraph{The near-horizon Rindler-like scattering problem}

The superscript ``Rindler'' in \eqref{eq:RindSmat} means that it is actually the scattering phase one would obtain by studying the scattering problem for the same free scalar but on the Rindler-like wedge
\begin{align}\label{eq:Rindlerlike}
	ds^2 = e^{-\frac{2}{\ell_\text{dS}}x} \left( -dt^2+dx^2 \right)+\ell_\text{dS}^2 d\Omega_{d-1}^2 \; , \qquad -\infty < x<\infty \; .
\end{align}
The Rindler horizon corresponds to $x =\infty$ while spatial infinity corresponds to $x=-\infty$. The scattering problem analogous to \eqref{eq:eff Sch} on this wedge is
\begin{align}\label{eq:Rindlerscatter}
	\left( -\partial_x^2 + M_l^2 e^{-\frac{2}{\ell_\text{dS}}x} \right) \psi(x) = \omega^2 \psi(x) \; , \qquad M_l \equiv \sqrt{\frac{l(l+d-2)}{\ell_\text{dS}}+m^2} \; .
\end{align}
Solving for the normalizable solution that exponentially decays at $x\to \infty$, one finds that the scattering phase is given by \eqref{eq:RindSmat}. 

Since the metric \eqref{eq:Rindlerlike} is nothing but the near-horizon limit of that the full static patch \eqref{eq:staticmetric}, the full S-matrix \eqref{eq:dSSmat} can be understood in the following way: it consists of a universal part \eqref{eq:RindSmat} that captures the near-horizon region, which knows nothing about the static patch geometry (except its temperature) and the scalar; the effect of the non-trivial information about the geometry, scalar mass and angular momentum is to ``dress" \eqref{eq:RindSmat} by an extra non-universal phase \eqref{eq:dSdress}.

\paragraph{Scattering phases and the spectral density}

Putting a brick wall cutoff \cite{tHooft:1984kcu} at some large value $x=R$ and imposing a Dirichlet boundary condition,\footnote{As emphasized in \cite{Law:2022zdq}, this choice of boundary condition turns out to be irrelevant. In particular, the result \eqref{eq:dosphase} only depends on the asymptotic behavior \eqref{eq:scattasym} of the scattering mode (and that in the reference system) but not its value on the brick wall $x=R$.} one can compute a smoothed density of states
\begin{align}
	\rho_l^R(\omega) = \frac{R+ \theta'_l (\omega)}{\pi}  +O\left(\frac{1}{R}\right) \; , \qquad \mathcal{S}_l (\omega) \equiv e^{2i\theta_l(\omega)}\; , 
\end{align}
whose divergence as $R\to \infty$ is the origin of the ill-definedness of the trace in \eqref{eq:PItrace}. However, if we compare our original problem \eqref{eq:eff Sch} with {\it some} reference problem, their {\it difference} in DOS is finite as $R\to \infty$:
\begin{align}\label{eq:dosphase}
	\Delta\rho_l (\omega) \equiv \rho_l(\omega)-\bar\rho_l (\omega) = \frac{1}{2\pi i} \partial_\omega \log \frac{\mathcal{S}_l (\omega)}{\bar{\mathcal{S}}_l (\omega)}
\end{align}
where $\bar\rho_l (\omega)$ and $\bar{\mathcal{S}}_l (\omega)$ are the DOS and S-matrix for the reference system. The relation \eqref{eq:dosphase} suggests that as opposed to the free energy \eqref{intro:log Z bulk} itself, {\it differences} in free energies are well-defined (up to the usual UV-divergence coming from integrating over all energies):
\begin{align}\label{eq:diffPF}
	\log Z_\text{bulk} - \log \bar Z_\text{bulk}= -\int_0^\infty d\omega \, \Delta \rho(\omega) \log \left( e^{\beta_\text{dS} \omega/2}- e^{-\beta_\text{dS} \omega /2}\right)  \; .
\end{align}
A priori, there is no unique choice for the reference system. For instance, we can choose it to be the one with zero potential (i.e. an empty box with size $R$), in which case the smoothed density $\bar\rho_l^R(\omega) =\frac{R}{\pi} $ is uniform and has a constant scattering phase. 

Uniquely fixing a reference system requires an extra physical input. In the current case, such an input is provided by demanding \eqref{eq:diffPF} to equal the Euclidean path integral \eqref{eq:scalarPI}. As it turns out, such a requirement fixes the reference system to be the scattering problem on the Rindler-like wedge \eqref{eq:Rindlerscatter}, i.e. we choose $\bar{\mathcal{S}}_l$ to be  \eqref{eq:RindSmat}. With this choice, 
\begin{gather}
	\Delta\rho^\text{dS} _l (\omega) = \frac{1}{2\pi i} \partial_\omega\log 	\mathcal{S}_l^\text{dS}  (\omega)\quad \; .
\end{gather}
Summing over $l\geq 0$, one then recovers the character-regularized DOS \eqref{eq:chardos}:
\begin{align}
	\Delta\rho^\text{dS} (\omega) =\sum_{l=0}^\infty D_l^d \Delta\rho^\text{dS}_l (\omega) = \int_{0}^\infty \frac{dt}{2\pi}\left( e^{i\omega t}+e^{-i\omega t} \right) \chi_{[\Delta, s=0]}(t)= \tilde{\rho}_{[\Delta, s=0]}(\omega) \;. 
\end{align}
Therefore, these considerations imply that the quasicanonical partition function \eqref{eq:canPF} (and therefore by \eqref{eq:PIPFs0} the Euclidean path integral \eqref{eq:scalarPI}) should be more accurately thought of as a {\it ratio}
\begin{align}\label{eq:Zbulkratio}
	\widetilde{Z}_\text{bulk} \equiv \frac{Z_\text{bulk}}{ Z^\text{Rindler}_\text{bulk}} \; ,
\end{align}
where $Z_\text{bulk}\equiv \text{``Tr"} \, e^{-\beta_\text{dS} \hat{H}}$ and $Z^\text{Rindler}_\text{bulk}\equiv \text{``Tr"} \, e^{-\beta_\text{dS} \hat{H}_\text{Rindler}}$ are the formally defined canonical partition functions for the scalar on the static patch and the Rindler-like wedge \eqref{eq:Rindlerlike} respectively, at the dS temperature $T_\text{dS}$.

%
%
%



\section{1-loop partition functions for spinning fields}\label{sec:PI}

In this section we discuss some of the qualitatively new features that arise as one extends our previous discussions to spinning fields. We will set the dS length (or sphere radius) $\ell_\text{dS}$ to 1.

\subsection{Spinning fields on $S^{d+1}$}

We first highlight the subtleties for their 1-loop path integrals on $S^{d+1}$, which have been studied in detail in \cite{Law:2020cpj}.

\subsubsection{Massive fields}

Let us start by looking at the example of a free massive vector, whose sphere path integral is
\begin{align}\label{eq:vecPI}
	Z^{(s=1,m^2)}_\text{PI}=\int \mathcal{D}A \, e^{-S[A]}\; , \qquad S[A]=\int_{S^{d+1}} \left( \frac{1}{4}F_{\mu\nu}F^{\mu\nu} +\frac{m^2}{2} A_\mu A^\mu\right) \; .
\end{align}
Here $S[A]$ is the Proca action. To proceed, one decomposes the vector $A$ into transverse and longitudinal components \cite{Babelon:1979wd,Mazur:1989by,Bern:1990bh}
\begin{align}\label{eq:vecde}
	A_\mu = A^T_\mu +\partial_\mu \phi\; , \qquad \nabla^\mu A^T_\mu  =0 \; .
\end{align}
Here is a subtlety that arises on any compact space: {\it to ensure this decomposition to be unique, one must exclude the normalizable constant mode of the longitudinal scalar $\phi$.} In other words, the path integration measure changes as
\begin{align}\label{eq:vecmeasure}
	\mathcal{D}A = J	\mathcal{D}A^T	\mathcal{D}'\phi  \; , 
\end{align}
where the prime denotes the omission of the constant mode, and $J$ is the Jacobian associated with the change of field variables $A\to \{A^T,\phi\}$. One then substitutes \eqref{eq:vecde}, \eqref{eq:vecmeasure} into \eqref{eq:vecPI} and finds
\begin{align}\label{eq:massvecdet}
	Z^{(s=1,m^2)}_\text{PI}=\det(-\nabla_{(1)}^2+m^2+d)^{-1/2} (m^2)^{1/2}\; .
\end{align}
Here the functional determinant is familiar and comes from the integration over $A^T$ and $\nabla_{(1)}^2$ is the Laplacian acting on transverse vector fields. The factor $(m^2)^{1/2}$, on the other hand, is only present for Euclidean path integrals on a compact space (in contrast to say $EAdS$), and arises from integrating over $\phi$ while keeping track of the omitted constant mode. 

This remains true for general spinning fields: the path integrations over their longitudinal components will lead to a finite number of factors analogous to the factor $(m^2)^{1/2}$. Such contributions must be present in order to be consistent with locality.\footnote{If such contributions are not included, one would find a non-zero logarithmic divergence even in odd dimensions.}

\subsubsection{Massless gauge fields}

Due to gauge invariance, the path integrals are more intricate for massless gauge fields.

\paragraph{The residual group volume}

Let us start with the example of a $U(1)$ gauge field, with path integral 
\begin{align}\label{eq:Max PI}
	Z^{U(1)}_\text{PI}= \frac{1}{\text{Vol}(\mathcal{G})} \int \mathcal{D}A \, e^{-S[A]}\,\, ,\qquad \text{Vol}(\mathcal{G})= \int \mathcal{D}\alpha=\prod_n  \frac{d\alpha_n }{\sqrt{2\pi}}\; .
\end{align}
The Maxwell action $S[A]=\int_{S^{d+1}} \frac{1}{4}F_{\mu\nu}F^{\mu\nu} $ is invariant under the gauge transformations 
\begin{align}\label{eq:gauge trans}
	A_\mu \to A_\mu +\partial_\mu \phi \; . 
\end{align}
The volume $\text{Vol}(\mathcal{G})$ of the space of gauge transformations is a path integration over a scalar field $\alpha$, the division by which in \eqref{eq:Max PI} compensates for the over-counting of gauge-equivalent orbits related by \eqref{eq:gauge trans}. Proceeding by doing the geometric decomposition \eqref{eq:vecde}, $\text{Vol}(\mathcal{G})$ would cancel against the integration $\int \mathcal{D}'\phi$ over the longitudinal scalar $\phi$. However, as pointed out in \cite{Donnelly:2013tia}, as opposed to $\int \mathcal{D}'\phi$, in $\text{Vol}(\mathcal{G})$ we {\it do not} exclude the constant mode $\alpha_0$ for consistency with locality and unitarity.\footnote{Again, one would find a non-zero logarithmic divergence in odd dimensions if this mode is not included.} This implies that the cancellation between $\text{Vol}(\mathcal{G})$ and $\int \mathcal{D}'\phi$ is over-complete, leaving a residual factor
\begin{align}\label{eq:group vol}
	\text{Vol}(U(1))_\text{PI}=\int \frac{d\alpha_0}{\sqrt{2\pi}} \; .
\end{align}
If we take into account the possible interactions of the Maxwell field $A_\mu$ with other matter fields, the mode $\alpha_0$ generates {\it global} $U(1)$ symmetries on the matter fields. \eqref{eq:group vol} is therefore the volume of the global $U(1)$  symmetry group. Taking this into account, one eventually arrives at
\begin{align}
	Z^{U(1)}_\text{PI}=Z_\text{G}Z_\text{Char} \; .
\end{align}
The factor
\begin{align}\label{eq:maxZch}
	Z_\text{Char}=\left(d(d-2)\right)^{\frac{1}{2}}\left(\frac{ \det'(-\nabla_{(0)}^2)}{\det(-\nabla_{(1)}^2 +d)}\right)^{\frac{1}{2}}
\end{align}
contains a ratio of determinants, where the denominator is the massless limit of the massive determinant \eqref{eq:massvecdet}; the numerator accounts for the quotient by the pure gauge modes, with the prime denotes the omission of the constant scalar mode. The subscript ``Char'' means that this part can be written in terms of the $SO(1,d+1)$ characters, which we will come back to in the next section.

The other factor
\begin{align}\label{eq:u1ZG}
	Z_\text{G}=\frac{1}{\text{Vol}(U(1))_\text{can}} \frac{\mathrm{g}}{\sqrt{(d-2)\text{Vol}(S^{d-1})}} 
\end{align}
is essentially \eqref{eq:group vol} modulo the numerical factors $\left(d(d-2)\right)^{\frac{1}{2}}$ in \eqref{eq:maxZch}. The dependence on the gauge coupling constant $\mathrm{g}$ comes from re-expressing the volume \eqref{eq:group vol}in terms of the volume $\text{Vol}(U(1))_\text{c}$ measured with respect to a ``canonical metric" which normalizes the $U(1)$ generator to unity. 


\paragraph{Conformal factor problem and the Polchinski's phase}

Another example is linearized Einstein gravity on $S^{d+1}$, which has a long and dramatic history \cite{Gibbons:1978ji, Christensen:1979iy,Fradkin:1983mq, Allen:1983dg,Taylor:1989ua,GRIFFIN1989295, Mazur:1989ch, Vassilevich:1992rk, Volkov:2000ih, Polchinski:1988ua}. A careful analysis brings the path integral into the form \cite{Law:2020cpj}
\begin{align}\label{eq:spin2PI}
	Z_\text{PI}=i^{-d-3}  Z_\text{G}Z_\text{Char} \; .
\end{align}
The factors $Z_\text{Char} $ and $Z_\text{G}$ are analogous to the $U(1)$ case. Specifically, 
\begin{align}
	Z_\text{Char}=&\bigg(d(d+2) \bigg)^{\frac{(d+1)(d+2)}{4}}  \frac{(d-1)^{\frac{d+2}{2}}}{(2d)^{1/2}}\frac{\det'(-\nabla_{(1)}^2-d)^{1/2}}{\det(-\nabla_{(2)}^2+2)^{1/2}} 
\end{align}
contains a ratio of determinants capturing the gravition fluctuations and the division by linearlized diffeomorphisms, while 
\begin{align}
	Z_G=\frac{\gamma^{\frac{(d+1)(d+2)}{2}}}{ \text{Vol}(SO(d+2))_\text{c}}\; ,\qquad  \gamma=\sqrt{\frac{8\pi G_N}{\text{Vol}(S^{d-1})}}
\end{align}
comes from integrating over the Killing vector modes generating the $SO(d+2)$ isometries. In particular, $Z_G$ depends on the Newton constant $G_N$ and the canonical isometry group volume $\text{Vol}(SO(d+2))_\text{c}$. Note that this is the origin for the logarithmic term in the quantum de Sitter entropy \eqref{eq:dsentropyex}.

Compared to the $U(1)$ case, there is a new contribution $i^{-d-3}$ in the path integral \eqref{eq:spin2PI}. To understand the origin of this factor, we note that in manipulating the path integral, we decompose the spin-2 field $h_{\mu\nu}$ into
\begin{align}\label{eq:spin 2 CoV}
		h_{\mu\nu}=h_{\mu\nu}^\text{TT} +\frac{1}{\sqrt{2}} (\nabla_{\mu} \xi_{\nu}+\nabla_{\nu} \xi_{\mu})+\frac{g_{\mu\nu}}{\sqrt{d+1}}\tilde{h}
\end{align} 
where $h_{\mu\nu}^\text{TT}$ is the transverse-traceless part of $h_{\mu\nu}$ satisfying $\nabla^\lambda h_{\lambda \mu}=0=h\indices{^\lambda_\lambda}$, $\xi_{\mu}$ the pure gauge part of $h_{\mu\nu}$, and $\tilde{h}$ the trace of $h_{\mu\nu}$. For \eqref{eq:spin 2 CoV} to be unique, we require $\xi_{\nu}$ to be orthogonal to all Killing vectors and $\tilde{h}$ to be orthogonal to all conformal Killing vectors on $S^{d+1}$ . Now, further expanding these components in terms of spherical harmonics on $S^{d+1}$, one finds that all but a finite number of modes in the trace part $\tilde{h}$ have a negative kinetic term \cite{Gibbons:1978ac}. A standard procedure \cite{Gibbons:1978ac} to cure this ``conformal factor problem" is to Wick-rotate $\tilde{h}\to  i \tilde{h}$ so that the path integral is well-defined. However, since a finite number of modes have a positive kinetic term to begin with, we must Wick-rotate them back, which eventually leads to this overall phase $i^{-d-3}$ first derived by Polchinski \cite{Polchinski:1988ua}. As shown in \cite{Law:2020cpj}, an analogous factor is present for any (partially) massless gauge fields with spin $s\geq 2$.

\subsection{The edge partition function}

Finally, let us make a connection with the Lorentzian/canonical picture discussed in Section \ref{sec:qnmPF} and \ref{sec:scattering}. To start with, one can still define a quasicanonical partition function $\widetilde{Z}_\text{bulk}$ by replacing in \eqref{eq:canPF} the character with its spinning counterpart. $\widetilde{Z}_\text{bulk}$ still has an interpretation \eqref{eq:Zbulkratio} as a ratio between the static patch and Rindler-like canonical partition functions.\footnote{To see this explicitly, one could repeat the analysis done in \cite{Grewal:2022hlo} for the case of massive higher-spin field on static BTZ or massive vector on Nariai.} However, it turns out $\widetilde{Z}_\text{bulk}$ is {\it not} equal to the Euclidean path integral $Z_\text{PI}$. Instead, $Z_\text{PI}$ receives some ``edge" contributions. 

For example, the 1-loop path integral for a massive spin-$s$ field takes the form
\begin{align} \label{intro formuhs2}
	\log Z_{\rm PI} =  \log \widetilde{Z}_{\rm bulk} - \log Z_{\rm edge} = \int_0^\infty \frac{dt}{2t} \, \frac{1+e^{-t}}{1-e^{-t}} \left( \chi_{\rm bulk}(t)- \chi_{\rm edge}(t) \right)   \; .
\end{align}
Here $\widetilde{Z}_{\rm bulk}$ is the quasicanonical partition function \eqref{eq:canPF} and we recall that the massive spin-$s$ character is given by \eqref{eq:masschar}. In $Z_{\rm edge}$, the edge character is explicitly given by
\begin{align} \label{intro chibulkedgeprev}
 \chi_{\rm edge} (t)=  D_{s-1}^{d+2} \, \frac{e^{-(\Delta-1)t}+e^{-(\bar \Delta-1)t}}{(1-e^{-t})^{d-2}}\;.
\end{align}

The case of massless gauge fields works similarly, except that there is an extra contribution from the global group volume factor (and the Polchinski's phase for spin $s\geq 2$). For instance, the $U(1)$ path integral takes the form $\log Z_{\rm PI} =  \log Z_\text{G}+\log Z_\text{Char} $, where $Z_\text{G}$ is given by \eqref{eq:u1ZG} and 
\begin{align}
	\log Z_\text{Char} = & \log \widetilde{Z}_{\rm bulk} - \log Z_{\rm edge}=\int_0^\infty \frac{dt}{2t} \, \frac{1+e^{-t}}{1-e^{-t}} \left( \chi_{\rm bulk}(t)- \chi_{\rm edge}(t) \right)
\end{align}
with bulk character \eqref{eq:masslessvecchar} and edge character
\begin{align}\label{eq:masslessvecedge}
	\chi_{\rm edge} (t) = 1+e^{-(d-2)t}-\frac{1}{(1-e^{-t})^{d-2}} \; .
\end{align}

Comparing \eqref{intro chibulkedgeprev} with \eqref{eq:masschar}, or \eqref{eq:masslessvecedge} with \eqref{eq:masslessvecchar}, we see that the edge characters live in two lower dimensions than their bulk counterparts. In other words, $Z_{\rm edge}$ is a path integral on a co-dimension-2 sphere, i.e. $S^{d-1}$. This suggests that these describe degrees of freedom living on the bifurcation surface of the dS horizon.


\section{Outlook}\label{sec:remarks}

While the search for a microscopic model for the dS horizon is far from complete, it could be informative to take a closer examination of the low-energy effective field theory. In this note, we have discussed the 1-loop Euclidean path integrals around the round sphere saddle, envisioning the prospect of constraining microscopic models. Conceptually, we have clarified their Lorentzian/canonical interpretation (at least for scalars), all thanks to  the powerful tools from $SO(1,d+1)$ representation theory. We would like to conclude with commenting on two intriguing future directions.

\paragraph{Algebra of observables}

The canonical partition function \eqref{eq:PItrace} can be viewed as the normalization of the reduced density matrix obtained by tracing out the global Bunch-Davis state along the antipodal static patch. This assumes that the global Hilbert space factorizes. From an algebraic QFT viewpoint (reviewed in \cite{Witten:2018zxz} and discussed in the context of static patch in \cite{Chandrasekaran:2022cip}), such a factorization does not really exist; the algebra of observables for a QFT in a static patch is a Type III von Neumann algebra, which does not admit a trace. The infinity of the single-particle DOS $\rho(\omega)$ in \eqref{intro:dosdis} is a manifestation of this non-existence of the trace and thus non-factorization of the global Hilbert space. It is then curious that a version of a trace and therefore  quasicanonical free energy \eqref{eq:canPF} can be defined using the $SO(1,d+1)$ Harish-Chandra character. Of course, this does not contradict the considerations in the previous paragraph: as discussed in  Section \ref{sec:scattering}, \eqref{eq:canPF} is not really a free energy but instead a ``renormalized" one. In any case, it would be very interesting to further understand the results presented in this article in terms of algebra of observables associated with a static patch.

\paragraph{Edge modes}

In section \ref{sec:PI}, we discussed a bulk-edge split for spinning path integrals, which certainly resonates with studies of entanglement entropy in gauge theories and gravity. In the early work \cite{Kabat:1995eq}, Kabat found a ``contact term'' in the conical entropy for Maxwell theory on black holes, which sparked an extensive investigation (see \cite{Solodukhin:2011gn,Zhitnitsky:2011tr,Donnelly:2011hn,Solodukhin:2012jh,Donnelly:2012st,Casini:2013rba,Donnelly:2014fua,Huang:2014pfa,Donnelly:2015hxa,Casini:2015dsg,Ghosh:2015iwa,Soni:2015yga,Soni:2016ogt,Donnelly:2016auv,Agarwal:2016cir,Blommaert:2018rsf,Blommaert:2018oue,Lin:2018bud,David:2022jfd} for a partial list) into the proper interpretation for such a contribution as coming from ``edge'' degrees of freedom living on the the bifurcation surface $S^{d-1}$. In our case of $S^{d+1}$ path integrals, while the physical meaning for the bulk part $Z_{\rm bulk}$ is clear, that for the edge part $Z_{\rm edge}$ is obscure at this point. With the inspirations from these past works, one might be able to clarify  the canonical picture for $Z_{\rm edge}$, and it would be very interesting to do so.


\section*{Acknowledgments}


\noindent This note is based on work in collaboration with Dionysios Anninos, Frederik Denef, Manvir Grewal, Klaas Parmentier, and Zimo Sun. I thank the organizers of the Workshop on  Features of a Quantum de Sitter Universe for the opportunity to present these results. I would like to also thank the workshop participants for fruitful discussions. This work  was supported in part by the Croucher Foundation and the Black Hole Initiative at Harvard University.


 \bibliographystyle{utphys}
\bibliography{BHChar}

\providecommand{\href}[2]{#2}\begingroup\raggedright\begin{thebibliography}{10}

\bibitem{PhysRevD.15.2738}
G.~W. Gibbons and S.~W. Hawking, ``Cosmological event horizons, thermodynamics,
  and particle creation,''
  \href{http://dx.doi.org/10.1103/PhysRevD.15.2738}{{\em Phys. Rev. D}
  {\bfseries 15} (May, 1977) 2738--2751}.
  \url{https://link.aps.org/doi/10.1103/PhysRevD.15.2738}.

\bibitem{Banerjee:2010qc}
S.~Banerjee, R.~K. Gupta, and A.~Sen, ``{Logarithmic Corrections to Extremal
  Black Hole Entropy from Quantum Entropy Function},''
  \href{http://dx.doi.org/10.1007/JHEP03(2011)147}{{\em JHEP} {\bfseries 03}
  (2011) 147}, \href{http://arxiv.org/abs/1005.3044}{{\ttfamily arXiv:1005.3044
  [hep-th]}}.

\bibitem{Banerjee:2011jp}
S.~Banerjee, R.~K. Gupta, I.~Mandal, and A.~Sen, ``{Logarithmic Corrections to
  N=4 and N=8 Black Hole Entropy: A One Loop Test of Quantum Gravity},''
  \href{http://dx.doi.org/10.1007/JHEP11(2011)143}{{\em JHEP} {\bfseries 11}
  (2011) 143}, \href{http://arxiv.org/abs/1106.0080}{{\ttfamily arXiv:1106.0080
  [hep-th]}}.

\bibitem{Sen:2012dw}
A.~Sen, ``{Logarithmic Corrections to Schwarzschild and Other Non-extremal
  Black Hole Entropy in Different Dimensions},''
  \href{http://dx.doi.org/10.1007/JHEP04(2013)156}{{\em JHEP} {\bfseries 04}
  (2013) 156}, \href{http://arxiv.org/abs/1205.0971}{{\ttfamily arXiv:1205.0971
  [hep-th]}}.

\bibitem{Sen:2012kpz}
A.~Sen, ``{Logarithmic Corrections to N=2 Black Hole Entropy: An Infrared
  Window into the Microstates},''
  \href{http://dx.doi.org/10.1007/s10714-012-1336-5}{{\em Gen. Rel. Grav.}
  {\bfseries 44} no.~5, (2012) 1207--1266},
  \href{http://arxiv.org/abs/1108.3842}{{\ttfamily arXiv:1108.3842 [hep-th]}}.

\bibitem{Sen:2014aja}
A.~Sen, ``{Microscopic and Macroscopic Entropy of Extremal Black Holes in
  String Theory},'' \href{http://dx.doi.org/10.1007/s10714-014-1711-5}{{\em
  Gen. Rel. Grav.} {\bfseries 46} (2014) 1711},
  \href{http://arxiv.org/abs/1402.0109}{{\ttfamily arXiv:1402.0109 [hep-th]}}.

\bibitem{Giombi:2013fka}
S.~Giombi and I.~R. Klebanov, ``{One Loop Tests of Higher Spin AdS/CFT},''
  \href{http://dx.doi.org/10.1007/JHEP12(2013)068}{{\em JHEP} {\bfseries 12}
  (2013) 068}, \href{http://arxiv.org/abs/1308.2337}{{\ttfamily arXiv:1308.2337
  [hep-th]}}.

\bibitem{Giombi:2014iua}
S.~Giombi, I.~R. Klebanov, and B.~R. Safdi, ``{Higher Spin AdS$_{d+1}$/CFT$_d$
  at One Loop},'' \href{http://dx.doi.org/10.1103/PhysRevD.89.084004}{{\em
  Phys. Rev. D} {\bfseries 89} no.~8, (2014) 084004},
  \href{http://arxiv.org/abs/1401.0825}{{\ttfamily arXiv:1401.0825 [hep-th]}}.

\bibitem{Giombi:2016pvg}
S.~Giombi, I.~R. Klebanov, and Z.~M. Tan, ``{The ABC of Higher-Spin AdS/CFT},''
  \href{http://dx.doi.org/10.3390/universe4010018}{{\em Universe} {\bfseries 4}
  no.~1, (2018) 18}, \href{http://arxiv.org/abs/1608.07611}{{\ttfamily
  arXiv:1608.07611 [hep-th]}}.

\bibitem{Gunaydin:2016amv}
M.~G\"unaydin, E.~D. Skvortsov, and T.~Tran, ``{Exceptional $F(4)$ higher-spin
  theory in AdS$_{6}$ at one-loop and other tests of duality},''
  \href{http://dx.doi.org/10.1007/JHEP11(2016)168}{{\em JHEP} {\bfseries 11}
  (2016) 168}, \href{http://arxiv.org/abs/1608.07582}{{\ttfamily
  arXiv:1608.07582 [hep-th]}}.

\bibitem{Anninos:2020hfj}
D.~Anninos, F.~Denef, Y.~T.~A. Law, and Z.~Sun, ``{Quantum de Sitter horizon
  entropy from quasicanonical bulk, edge, sphere and topological string
  partition functions},'' \href{http://dx.doi.org/10.1007/JHEP01(2022)088}{{\em
  JHEP} {\bfseries 01} (2022) 088},
  \href{http://arxiv.org/abs/2009.12464}{{\ttfamily arXiv:2009.12464
  [hep-th]}}.

\bibitem{Shyam:2021ciy}
V.~Shyam, ``{$ \mathrm{T}\overline{\mathrm{T}} $ + \ensuremath{\Lambda}$_{2}$
  deformed CFT on the stretched dS$_{3}$ horizon},''
  \href{http://dx.doi.org/10.1007/JHEP04(2022)052}{{\em JHEP} {\bfseries 04}
  (2022) 052}, \href{http://arxiv.org/abs/2106.10227}{{\ttfamily
  arXiv:2106.10227 [hep-th]}}.

\bibitem{Coleman:2021nor}
E.~Coleman, E.~A. Mazenc, V.~Shyam, E.~Silverstein, R.~M. Soni, G.~Torroba, and
  S.~Yang, ``{De Sitter microstates from T$ \overline{T} $ +
  \ensuremath{\Lambda}$_{2}$ and the Hawking-Page transition},''
  \href{http://dx.doi.org/10.1007/JHEP07(2022)140}{{\em JHEP} {\bfseries 07}
  (2022) 140}, \href{http://arxiv.org/abs/2110.14670}{{\ttfamily
  arXiv:2110.14670 [hep-th]}}.

\bibitem{Law:2020cpj}
Y.~T.~A. Law, ``{A compendium of sphere path integrals},''
  \href{http://dx.doi.org/10.1007/JHEP12(2021)213}{{\em JHEP} {\bfseries 12}
  (2021) 213}, \href{http://arxiv.org/abs/2012.06345}{{\ttfamily
  arXiv:2012.06345 [hep-th]}}.

\bibitem{Law:2022zdq}
Y.~T.~A. Law and K.~Parmentier, ``{Black hole scattering and partition
  functions},'' \href{http://dx.doi.org/10.1007/JHEP10(2022)039}{{\em JHEP}
  {\bfseries 10} (2022) 039}, \href{http://arxiv.org/abs/2207.07024}{{\ttfamily
  arXiv:2207.07024 [hep-th]}}.

\bibitem{PhysRevD.15.2752}
G.~W. Gibbons and S.~W. Hawking, ``Action integrals and partition functions in
  quantum gravity,'' \href{http://dx.doi.org/10.1103/PhysRevD.15.2752}{{\em
  Phys. Rev. D} {\bfseries 15} (May, 1977) 2752--2756}.
  \url{https://link.aps.org/doi/10.1103/PhysRevD.15.2752}.

\bibitem{Hawking:1976ja}
S.~W. Hawking, ``{Zeta Function Regularization of Path Integrals in Curved
  Space-Time},'' \href{http://dx.doi.org/10.1007/BF01626516}{{\em Commun. Math.
  Phys.} {\bfseries 55} (1977) 133}.

\bibitem{Vassilevich:2003xt}
D.~V. Vassilevich, ``{Heat kernel expansion: User's manual},''
  \href{http://dx.doi.org/10.1016/j.physrep.2003.09.002}{{\em Phys. Rept.}
  {\bfseries 388} (2003) 279--360},
  \href{http://arxiv.org/abs/hep-th/0306138}{{\ttfamily arXiv:hep-th/0306138}}.

\bibitem{Sun:2021thf}
Z.~Sun, ``{A note on the representations of $\text{SO}(1,d+1)$},''
  \href{http://arxiv.org/abs/2111.04591}{{\ttfamily arXiv:2111.04591
  [hep-th]}}.

\bibitem{Higuchi:1986py}
A.~Higuchi, ``{Forbidden Mass Range for Spin-2 Field Theory in De Sitter
  Space-time},'' \href{http://dx.doi.org/10.1016/0550-3213(87)90691-2}{{\em
  Nucl. Phys. B} {\bfseries 282} (1987) 397--436}.

\bibitem{Zinoviev:2001dt}
Y.~M. Zinoviev, ``{On massive high spin particles in AdS},''
  \href{http://arxiv.org/abs/hep-th/0108192}{{\ttfamily arXiv:hep-th/0108192}}.

\bibitem{Deser:1983tm}
S.~Deser and R.~I. Nepomechie, ``{Anomalous Propagation of Gauge Fields in
  Conformally Flat Spaces},''
  \href{http://dx.doi.org/10.1016/0370-2693(83)90317-9}{{\em Phys. Lett. B}
  {\bfseries 132} (1983) 321--324}.

\bibitem{DESER1984396}
S.~Deser and R.~I. Nepomechie, ``Gauge invariance versus masslessness in de
  sitter spaces,''
  \href{http://dx.doi.org/https://doi.org/10.1016/0003-4916(84)90156-8}{{\em
  Annals of Physics} {\bfseries 154} no.~2, (1984) 396--420}.
  \url{https://www.sciencedirect.com/science/article/pii/0003491684901568}.

\bibitem{Brink:2000ag}
L.~Brink, R.~R. Metsaev, and M.~A. Vasiliev, ``{How massless are massless
  fields in AdS(d)},''
  \href{http://dx.doi.org/10.1016/S0550-3213(00)00402-8}{{\em Nucl. Phys. B}
  {\bfseries 586} (2000) 183--205},
  \href{http://arxiv.org/abs/hep-th/0005136}{{\ttfamily arXiv:hep-th/0005136}}.

\bibitem{Deser:2001pe}
S.~Deser and A.~Waldron, ``{Gauge invariances and phases of massive higher
  spins in (A)dS},''
  \href{http://dx.doi.org/10.1103/PhysRevLett.87.031601}{{\em Phys. Rev. Lett.}
  {\bfseries 87} (2001) 031601},
  \href{http://arxiv.org/abs/hep-th/0102166}{{\ttfamily arXiv:hep-th/0102166}}.

\bibitem{Deser:2001us}
S.~Deser and A.~Waldron, ``{Partial masslessness of higher spins in (A)dS},''
  \href{http://dx.doi.org/10.1016/S0550-3213(01)00212-7}{{\em Nucl. Phys. B}
  {\bfseries 607} (2001) 577--604},
  \href{http://arxiv.org/abs/hep-th/0103198}{{\ttfamily arXiv:hep-th/0103198}}.

\bibitem{Deser:2001wx}
S.~Deser and A.~Waldron, ``{Stability of massive cosmological gravitons},''
  \href{http://dx.doi.org/10.1016/S0370-2693(01)00523-8}{{\em Phys. Lett. B}
  {\bfseries 508} (2001) 347--353},
  \href{http://arxiv.org/abs/hep-th/0103255}{{\ttfamily arXiv:hep-th/0103255}}.

\bibitem{Deser:2001xr}
S.~Deser and A.~Waldron, ``{Null propagation of partially massless higher spins
  in (A)dS and cosmological constant speculations},''
  \href{http://dx.doi.org/10.1016/S0370-2693(01)00756-0}{{\em Phys. Lett. B}
  {\bfseries 513} (2001) 137--141},
  \href{http://arxiv.org/abs/hep-th/0105181}{{\ttfamily arXiv:hep-th/0105181}}.

\bibitem{Skvortsov:2006at}
E.~D. Skvortsov and M.~A. Vasiliev, ``{Geometric formulation for partially
  massless fields},''
  \href{http://dx.doi.org/10.1016/j.nuclphysb.2006.06.019}{{\em Nucl. Phys. B}
  {\bfseries 756} (2006) 117--147},
  \href{http://arxiv.org/abs/hep-th/0601095}{{\ttfamily arXiv:hep-th/0601095}}.

\bibitem{Hinterbichler:2016fgl}
K.~Hinterbichler and A.~Joyce, ``{Manifest Duality for Partially Massless
  Higher Spins},'' \href{http://dx.doi.org/10.1007/JHEP09(2016)141}{{\em JHEP}
  {\bfseries 09} (2016) 141}, \href{http://arxiv.org/abs/1608.04385}{{\ttfamily
  arXiv:1608.04385 [hep-th]}}.

\bibitem{Basile:2016aen}
T.~Basile, X.~Bekaert, and N.~Boulanger, ``{Mixed-symmetry fields in de Sitter
  space: a group theoretical glance},''
  \href{http://dx.doi.org/10.1007/JHEP05(2017)081}{{\em JHEP} {\bfseries 05}
  (2017) 081}, \href{http://arxiv.org/abs/1612.08166}{{\ttfamily
  arXiv:1612.08166 [hep-th]}}.

\bibitem{Bonifacio:2018zex}
J.~Bonifacio, K.~Hinterbichler, A.~Joyce, and R.~A. Rosen, ``{Shift Symmetries
  in (Anti) de Sitter Space},''
  \href{http://dx.doi.org/10.1007/JHEP02(2019)178}{{\em JHEP} {\bfseries 02}
  (2019) 178}, \href{http://arxiv.org/abs/1812.08167}{{\ttfamily
  arXiv:1812.08167 [hep-th]}}.

\bibitem{Brady:1999wd}
P.~R. Brady, C.~M. Chambers, W.~G. Laarakkers, and E.~Poisson, ``{Radiative
  falloff in Schwarzschild-de Sitter space-time},''
  \href{http://dx.doi.org/10.1103/PhysRevD.60.064003}{{\em Phys. Rev. D}
  {\bfseries 60} (1999) 064003},
  \href{http://arxiv.org/abs/gr-qc/9902010}{{\ttfamily arXiv:gr-qc/9902010}}.

\bibitem{Lopez-Ortega:2006aal}
A.~Lopez-Ortega, ``{Quasinormal modes of D-dimensional de Sitter spacetime},''
  \href{http://dx.doi.org/10.1007/s10714-006-0335-9}{{\em Gen. Rel. Grav.}
  {\bfseries 38} (2006) 1565--1591},
  \href{http://arxiv.org/abs/gr-qc/0605027}{{\ttfamily arXiv:gr-qc/0605027}}.

\bibitem{Lopez-Ortega:2012xvr}
A.~Lopez-Ortega, ``{On the quasinormal modes of the de Sitter spacetime},''
  \href{http://dx.doi.org/10.1007/s10714-012-1398-4}{{\em Gen. Rel. Grav.}
  {\bfseries 44} (2012) 2387--2400},
  \href{http://arxiv.org/abs/1207.6791}{{\ttfamily arXiv:1207.6791 [gr-qc]}}.

\bibitem{Sun:2020sgn}
Z.~Sun, ``{Higher spin de Sitter quasinormal modes},''
  \href{http://arxiv.org/abs/2010.09684}{{\ttfamily arXiv:2010.09684
  [hep-th]}}.

\bibitem{Ng:2012xp}
G.~S. Ng and A.~Strominger, ``{State/Operator Correspondence in Higher-Spin
  dS/CFT},'' \href{http://dx.doi.org/10.1088/0264-9381/30/10/104002}{{\em
  Class. Quant. Grav.} {\bfseries 30} (2013) 104002},
  \href{http://arxiv.org/abs/1204.1057}{{\ttfamily arXiv:1204.1057 [hep-th]}}.

\bibitem{Jafferis:2013qia}
D.~L. Jafferis, A.~Lupsasca, V.~Lysov, G.~S. Ng, and A.~Strominger,
  ``{Quasinormal quantization in de Sitter spacetime},''
  \href{http://dx.doi.org/10.1007/JHEP01(2015)004}{{\em JHEP} {\bfseries 01}
  (2015) 004}, \href{http://arxiv.org/abs/1305.5523}{{\ttfamily arXiv:1305.5523
  [hep-th]}}.

\bibitem{Tanhayi:2014kba}
M.~R. Tanhayi, ``{Quasinormal modes in de Sitter space: Plane wave method},''
  \href{http://dx.doi.org/10.1103/PhysRevD.90.064010}{{\em Phys. Rev. D}
  {\bfseries 90} no.~6, (2014) 064010},
  \href{http://arxiv.org/abs/1402.2893}{{\ttfamily arXiv:1402.2893 [gr-qc]}}.

\bibitem{tHooft:1984kcu}
G.~'t~Hooft, ``{On the Quantum Structure of a Black Hole},''
  \href{http://dx.doi.org/10.1016/0550-3213(85)90418-3}{{\em Nucl. Phys. B}
  {\bfseries 256} (1985) 727--745}.

\bibitem{Babelon:1979wd}
O.~Babelon and C.~M. Viallet, ``{The Geometrical Interpretation of the
  {Faddeev-Popov} Determinant},''
  \href{http://dx.doi.org/10.1016/0370-2693(79)90589-6}{{\em Phys. Lett. B}
  {\bfseries 85} (1979) 246--248}.

\bibitem{Mazur:1989by}
P.~O. Mazur and E.~Mottola, ``{The Gravitational Measure, Solution of the
  Conformal Factor Problem and Stability of the Ground State of Quantum
  Gravity},'' \href{http://dx.doi.org/10.1016/0550-3213(90)90268-I}{{\em Nucl.
  Phys. B} {\bfseries 341} (1990) 187--212}.

\bibitem{Bern:1990bh}
Z.~Bern, E.~Mottola, and S.~K. Blau, ``{General covariance of the path integral
  for quantum gravity},''
  \href{http://dx.doi.org/10.1103/PhysRevD.43.1212}{{\em Phys. Rev. D}
  {\bfseries 43} (1991) 1212--1222}.

\bibitem{Donnelly:2013tia}
W.~Donnelly and A.~C. Wall, ``{Unitarity of Maxwell theory on curved spacetimes
  in the covariant formalism},''
  \href{http://dx.doi.org/10.1103/PhysRevD.87.125033}{{\em Phys. Rev. D}
  {\bfseries 87} no.~12, (2013) 125033},
  \href{http://arxiv.org/abs/1303.1885}{{\ttfamily arXiv:1303.1885 [hep-th]}}.

\bibitem{Gibbons:1978ji}
G.~W. Gibbons and M.~J. Perry, ``{Quantizing Gravitational Instantons},''
  \href{http://dx.doi.org/10.1016/0550-3213(78)90434-0}{{\em Nucl. Phys. B}
  {\bfseries 146} (1978) 90--108}.

\bibitem{Christensen:1979iy}
S.~M. Christensen and M.~J. Duff, ``{Quantizing Gravity with a Cosmological
  Constant},'' \href{http://dx.doi.org/10.1016/0550-3213(80)90423-X}{{\em Nucl.
  Phys. B} {\bfseries 170} (1980) 480--506}.

\bibitem{Fradkin:1983mq}
E.~S. Fradkin and A.~A. Tseytlin, ``{One Loop Effective Potential in Gauged
  O(4) Supergravity},''
  \href{http://dx.doi.org/10.1016/0550-3213(84)90074-9}{{\em Nucl. Phys. B}
  {\bfseries 234} (1984) 472}.

\bibitem{Allen:1983dg}
B.~Allen, ``{Phase Transitions in de Sitter Space},''
  \href{http://dx.doi.org/10.1016/0550-3213(83)90470-4}{{\em Nucl. Phys. B}
  {\bfseries 226} (1983) 228--252}.

\bibitem{Taylor:1989ua}
T.~R. Taylor and G.~Veneziano, ``{Quantum Gravity at Large Distances and the
  Cosmological Constant},''
  \href{http://dx.doi.org/10.1016/0550-3213(90)90615-K}{{\em Nucl. Phys. B}
  {\bfseries 345} (1990) 210--230}.

\bibitem{GRIFFIN1989295}
P.~A. Griffin and D.~A. Kosower, ``Curved spacetime one-loop gravity in a
  physical gauge,''
  \href{http://dx.doi.org/https://doi.org/10.1016/0370-2693(89)91313-0}{{\em
  Physics Letters B} {\bfseries 233} no.~3, (1989) 295--300}.
  \url{https://www.sciencedirect.com/science/article/pii/0370269389913130}.

\bibitem{Mazur:1989ch}
P.~O. Mazur and E.~Mottola, ``{ABSENCE OF PHASE IN THE SUM OVER SPHERES},''.

\bibitem{Vassilevich:1992rk}
D.~V. Vassilevich, ``{One loop quantum gravity on de Sitter space},''
  \href{http://dx.doi.org/10.1142/S0217751X93000679}{{\em Int. J. Mod. Phys. A}
  {\bfseries 8} (1993) 1637--1652}.

\bibitem{Volkov:2000ih}
M.~S. Volkov and A.~Wipf, ``{Black hole pair creation in de Sitter space: A
  Complete one loop analysis},''
  \href{http://dx.doi.org/10.1016/S0550-3213(00)00287-X}{{\em Nucl. Phys. B}
  {\bfseries 582} (2000) 313--362},
  \href{http://arxiv.org/abs/hep-th/0003081}{{\ttfamily arXiv:hep-th/0003081}}.

\bibitem{Polchinski:1988ua}
J.~Polchinski, ``{The Phase of the Sum Over Spheres},''
  \href{http://dx.doi.org/10.1016/0370-2693(89)90387-0}{{\em Phys. Lett. B}
  {\bfseries 219} (1989) 251--257}.

\bibitem{Gibbons:1978ac}
G.~W. Gibbons, S.~W. Hawking, and M.~J. Perry, ``{Path Integrals and the
  Indefiniteness of the Gravitational Action},''
  \href{http://dx.doi.org/10.1016/0550-3213(78)90161-X}{{\em Nucl. Phys. B}
  {\bfseries 138} (1978) 141--150}.

\bibitem{Grewal:2022hlo}
M.~Grewal, Y.~T.~A. Law, and K.~Parmentier, ``{Black Hole Horizon Edge
  Partition Functions},'' \href{http://arxiv.org/abs/2211.16644}{{\ttfamily
  arXiv:2211.16644 [hep-th]}}.

\bibitem{Witten:2018zxz}
E.~Witten, ``{APS Medal for Exceptional Achievement in Research: Invited
  article on entanglement properties of quantum field theory},''
  \href{http://dx.doi.org/10.1103/RevModPhys.90.045003}{{\em Rev. Mod. Phys.}
  {\bfseries 90} no.~4, (2018) 045003},
  \href{http://arxiv.org/abs/1803.04993}{{\ttfamily arXiv:1803.04993
  [hep-th]}}.

\bibitem{Chandrasekaran:2022cip}
V.~Chandrasekaran, R.~Longo, G.~Penington, and E.~Witten, ``{An Algebra of
  Observables for de Sitter Space},''
  \href{http://arxiv.org/abs/2206.10780}{{\ttfamily arXiv:2206.10780
  [hep-th]}}.

\bibitem{Kabat:1995eq}
D.~N. Kabat, ``{Black hole entropy and entropy of entanglement},''
  \href{http://dx.doi.org/10.1016/0550-3213(95)00443-V}{{\em Nucl. Phys. B}
  {\bfseries 453} (1995) 281--299},
  \href{http://arxiv.org/abs/hep-th/9503016}{{\ttfamily arXiv:hep-th/9503016}}.

\bibitem{Solodukhin:2011gn}
S.~N. Solodukhin, ``{Entanglement entropy of black holes},''
  \href{http://dx.doi.org/10.12942/lrr-2011-8}{{\em Living Rev. Rel.}
  {\bfseries 14} (2011) 8}, \href{http://arxiv.org/abs/1104.3712}{{\ttfamily
  arXiv:1104.3712 [hep-th]}}.

\bibitem{Zhitnitsky:2011tr}
A.~R. Zhitnitsky, ``{Entropy, Contact Interaction with Horizon and Dark
  Energy},'' \href{http://dx.doi.org/10.1103/PhysRevD.84.124008}{{\em Phys.
  Rev. D} {\bfseries 84} (2011) 124008},
  \href{http://arxiv.org/abs/1105.6088}{{\ttfamily arXiv:1105.6088 [hep-th]}}.

\bibitem{Donnelly:2011hn}
W.~Donnelly, ``{Decomposition of entanglement entropy in lattice gauge
  theory},'' \href{http://dx.doi.org/10.1103/PhysRevD.85.085004}{{\em Phys.
  Rev. D} {\bfseries 85} (2012) 085004},
  \href{http://arxiv.org/abs/1109.0036}{{\ttfamily arXiv:1109.0036 [hep-th]}}.

\bibitem{Solodukhin:2012jh}
S.~N. Solodukhin, ``{Remarks on effective action and entanglement entropy of
  Maxwell field in generic gauge},''
  \href{http://dx.doi.org/10.1007/JHEP12(2012)036}{{\em JHEP} {\bfseries 12}
  (2012) 036}, \href{http://arxiv.org/abs/1209.2677}{{\ttfamily arXiv:1209.2677
  [hep-th]}}.

\bibitem{Donnelly:2012st}
W.~Donnelly and A.~C. Wall, ``{Do gauge fields really contribute negatively to
  black hole entropy?},''
  \href{http://dx.doi.org/10.1103/PhysRevD.86.064042}{{\em Phys. Rev. D}
  {\bfseries 86} (2012) 064042},
  \href{http://arxiv.org/abs/1206.5831}{{\ttfamily arXiv:1206.5831 [hep-th]}}.

\bibitem{Casini:2013rba}
H.~Casini, M.~Huerta, and J.~A. Rosabal, ``{Remarks on entanglement entropy for
  gauge fields},'' \href{http://dx.doi.org/10.1103/PhysRevD.89.085012}{{\em
  Phys. Rev. D} {\bfseries 89} no.~8, (2014) 085012},
  \href{http://arxiv.org/abs/1312.1183}{{\ttfamily arXiv:1312.1183 [hep-th]}}.

\bibitem{Donnelly:2014fua}
W.~Donnelly and A.~C. Wall, ``{Entanglement entropy of electromagnetic edge
  modes},'' \href{http://dx.doi.org/10.1103/PhysRevLett.114.111603}{{\em Phys.
  Rev. Lett.} {\bfseries 114} no.~11, (2015) 111603},
  \href{http://arxiv.org/abs/1412.1895}{{\ttfamily arXiv:1412.1895 [hep-th]}}.

\bibitem{Huang:2014pfa}
K.-W. Huang, ``{Central Charge and Entangled Gauge Fields},''
  \href{http://dx.doi.org/10.1103/PhysRevD.92.025010}{{\em Phys. Rev. D}
  {\bfseries 92} no.~2, (2015) 025010},
  \href{http://arxiv.org/abs/1412.2730}{{\ttfamily arXiv:1412.2730 [hep-th]}}.

\bibitem{Donnelly:2015hxa}
W.~Donnelly and A.~C. Wall, ``{Geometric entropy and edge modes of the
  electromagnetic field},''
  \href{http://dx.doi.org/10.1103/PhysRevD.94.104053}{{\em Phys. Rev. D}
  {\bfseries 94} no.~10, (2016) 104053},
  \href{http://arxiv.org/abs/1506.05792}{{\ttfamily arXiv:1506.05792
  [hep-th]}}.

\bibitem{Casini:2015dsg}
H.~Casini and M.~Huerta, ``{Entanglement entropy of a Maxwell field on the
  sphere},'' \href{http://dx.doi.org/10.1103/PhysRevD.93.105031}{{\em Phys.
  Rev. D} {\bfseries 93} no.~10, (2016) 105031},
  \href{http://arxiv.org/abs/1512.06182}{{\ttfamily arXiv:1512.06182
  [hep-th]}}.

\bibitem{Ghosh:2015iwa}
S.~Ghosh, R.~M. Soni, and S.~P. Trivedi, ``{On The Entanglement Entropy For
  Gauge Theories},'' \href{http://dx.doi.org/10.1007/JHEP09(2015)069}{{\em
  JHEP} {\bfseries 09} (2015) 069},
  \href{http://arxiv.org/abs/1501.02593}{{\ttfamily arXiv:1501.02593
  [hep-th]}}.

\bibitem{Soni:2015yga}
R.~M. Soni and S.~P. Trivedi, ``{Aspects of Entanglement Entropy for Gauge
  Theories},'' \href{http://dx.doi.org/10.1007/JHEP01(2016)136}{{\em JHEP}
  {\bfseries 01} (2016) 136}, \href{http://arxiv.org/abs/1510.07455}{{\ttfamily
  arXiv:1510.07455 [hep-th]}}.

\bibitem{Soni:2016ogt}
R.~M. Soni and S.~P. Trivedi, ``{Entanglement entropy in (3 + 1)-d free U(1)
  gauge theory},'' \href{http://dx.doi.org/10.1007/JHEP02(2017)101}{{\em JHEP}
  {\bfseries 02} (2017) 101}, \href{http://arxiv.org/abs/1608.00353}{{\ttfamily
  arXiv:1608.00353 [hep-th]}}.

\bibitem{Donnelly:2016auv}
W.~Donnelly and L.~Freidel, ``{Local subsystems in gauge theory and gravity},''
  \href{http://dx.doi.org/10.1007/JHEP09(2016)102}{{\em JHEP} {\bfseries 09}
  (2016) 102}, \href{http://arxiv.org/abs/1601.04744}{{\ttfamily
  arXiv:1601.04744 [hep-th]}}.

\bibitem{Agarwal:2016cir}
A.~Agarwal, D.~Karabali, and V.~P. Nair, ``{Gauge-invariant Variables and
  Entanglement Entropy},''
  \href{http://dx.doi.org/10.1103/PhysRevD.96.125008}{{\em Phys. Rev. D}
  {\bfseries 96} no.~12, (2017) 125008},
  \href{http://arxiv.org/abs/1701.00014}{{\ttfamily arXiv:1701.00014
  [hep-th]}}.

\bibitem{Blommaert:2018rsf}
A.~Blommaert, T.~G. Mertens, H.~Verschelde, and V.~I. Zakharov, ``{Edge State
  Quantization: Vector Fields in Rindler},''
  \href{http://dx.doi.org/10.1007/JHEP08(2018)196}{{\em JHEP} {\bfseries 08}
  (2018) 196}, \href{http://arxiv.org/abs/1801.09910}{{\ttfamily
  arXiv:1801.09910 [hep-th]}}.

\bibitem{Blommaert:2018oue}
A.~Blommaert, T.~G. Mertens, and H.~Verschelde, ``{Edge dynamics from the path
  integral \textemdash{} Maxwell and Yang-Mills},''
  \href{http://dx.doi.org/10.1007/JHEP11(2018)080}{{\em JHEP} {\bfseries 11}
  (2018) 080}, \href{http://arxiv.org/abs/1804.07585}{{\ttfamily
  arXiv:1804.07585 [hep-th]}}.

\bibitem{Lin:2018bud}
J.~Lin and D.~Radi\v{c}evi\'c, ``{Comments on defining entanglement entropy},''
  \href{http://dx.doi.org/10.1016/j.nuclphysb.2020.115118}{{\em Nucl. Phys. B}
  {\bfseries 958} (2020) 115118},
  \href{http://arxiv.org/abs/1808.05939}{{\ttfamily arXiv:1808.05939
  [hep-th]}}.

\bibitem{David:2022jfd}
J.~R. David and J.~Mukherjee, ``{Entanglement entropy of gravitational edge
  modes},'' \href{http://dx.doi.org/10.1007/JHEP08(2022)065}{{\em JHEP}
  {\bfseries 08} (2022) 065}, \href{http://arxiv.org/abs/2201.06043}{{\ttfamily
  arXiv:2201.06043 [hep-th]}}.

\end{thebibliography}\endgroup

\end{document}